\documentclass[apj]{emulateapj}

\usepackage{amsmath,amssymb}
\usepackage{graphicx}

\slugcomment{Published in \apj.}
\received{Sept. 1, 2011}
\accepted{Dec. 12, 2012}
\published{Feb. 7, 2013}

\shorttitle{COOLING OF COLOR SUPERCONDUCTING COMPACT STARS}
\shortauthors{Noda et al.}

\begin{document}

\title{COOLING OF COMPACT STARS WITH COLOR SUPERCONDUCTING PHASE IN QUARK HADRON MIXED PHASE}

\author{Tsuneo Noda, Masa-aki Hashimoto}
\affil{Department of Physics, Kyushu University, 6-10-1 Hakozaki, Higashi-ku, Fukuoka, 812-8581 Japan; tsune@phys.kyushu-u.ac.jp, hashimoto@phys.kyushu-u.ac.jp}
\author{Nobutoshi Yasutake}
\affil{Research Center for the Early Universe, University of Tokyo, 7-3-1 Hongo, Bunkyo-ku, Tokyo, 113-0033 Japan}
\author{Toshiki Maruyama}
\affil{Advanced Science Research Center, Japan Atomic Energy Agency, Tokai, Ibaraki 319-1195 Japan}
\author{Toshitaka Tatsumi}
\affil{Department of Physics, Kyoto University, Kitashirakawa-Oiwake-cho, Kyoto 606-8502 Japan}
\and
\author{Masayuki Y. Fujimoto}
\affil{Department of Physics, Hokkaido University, Kita-10 Nishi-8, Kita-ku, Sapporo, Hokkaido 060-0810 Japan}

\begin{abstract}

We present a new scenario for the cooling of compact stars considering the central source of Cassiopeia A (Cas A). The Cas A observation shows that the central source is a compact star that has high effective temperature, and it is consistent with the cooling without exotic phases. The observation also gives the mass range of $M \geqslant 1.5 M_\odot$, which may conflict with the current plausible cooling scenario of compact stars. There are some cooled compact stars such as Vela or 3C58, which can be barely explained by the minimal cooling scenario, which includes the neutrino emission by nucleon superfluidity (PBF). Therefore, we invoke the exotic cooling processes, where a heavier star cools faster than lighter one. However, the scenario seems to be inconsistent with the observation of Cas A. Therefore, we present a new cooling scenario to explain the observation of Cas A by constructing models that include a quark color superconducting (CSC) phase with a large energy gap; this phase appears at ultrahigh density region and reduces neutrino emissivity. In our model, a compact star has CSC quark core with a low neutrino emissivity surrounded by high emissivity region made by normal quarks. We present cooling curves obtained from the evolutionary calculations of compact stars: while heavier stars cool slowly, and lighter ones indicate the opposite tendency without considering nucleon superfluidity. Furthermore, we show that our scenario is consistent with the recent observations of the effective temperature of Cas A during the last 10 years, including nucleon superfluidity.

\end{abstract}

\section{Introduction}

The cooling of compact stars has been discussed mainly in the context of neutron stars
for decades~\citep{st98,bk09}.
It has been believed that some stars require exotic cooling to explain the observed effective temperature and others can be explained by the modified URCA and Bremsstrahlung processes, where the central density of the star determines which cooling process works: an exotic cooling phase appears at higher density above a threshold density~\citep[e.g.,][]{yak05,gus05}.
As a consequence, the heavier star which has higher central density cools faster than lighter one~\citep{lat91}.
However, as described below this scenario becomes inconsistent when we consider the recent observation of the effective temperature of Cas A whose mass has been found to be unexpectedly large.

Cas A is the youngest-known supernova remnant in the Milky Way and it is located $\sim3.4~\mathrm{kpc}$ from the solar system~\citep{rd95}. The supernova explosion occurred about 330 years ago, but due to absorption by the interstellar medium, there are no exact historical records except for 
an unclear detection by J. Flamsteed in 1680~\citep{aw80}. Recently,
 \citet{ho09}  and \cite{ho10} have analyzed the X-ray spectra of Cas A. They
give the effective temperature
and possible regions occupied by mass and radius relations. Since Cas A is the isolated remnant, the uncertainty of mass--radius relation could be large.
The lowest mass obtained from the $\chi^2$ fitting is about $1.5M_\odot$. Considering the age of $t=330$~yr, $T_{\rm eff}$ of Cas A must occupy a point of a cooling curve
due to the scenario with modified URCA and Bremsstrahlung processes included on the $(T_{\rm eff}-t)$ plane. This gives strong constraint on the equation of state (EoS) and cooling processes. 
Furthermore, \citet{ho10} reported the observation of $T_\mathrm{eff}$ for Cas A in the past 10 years. \citet{yak11}, \citet{pg11}, and \citet{sht11} insist that the rapid decrease in $T_\mathrm{eff}$ over time shows that the transition to nucleon superfluidity occurs.

On the other hand, there are some cooled stars whose effective temperature cannot be explained
by the neutrino emission processes without nucleon superfluidity, including the modified URCA and Bremsstrahlung.
It needs stronger cooling process as in the case of J0205+6449 in 3C58 (hereafter ``3C58'') or Vela pulsar (B0833--45). Also an accreting neutron star SAX J1808 requires strong cooling. 3C58 and Vela may be explained by the minimal cooling model which includes nucleon superfluidity~\citep{pg09}.
However, SAX J1808 needs stronger cooling than the minimal cooling \citep{heinke08}.
If we consider the strong cooling process according to the conventional scenario, their masses should become larger than that of Cas A, which may be inconsistent with the mass observations of double neutron stars; the mass of each neutron star is nearly $M \sim1.4~M_\odot$ \citep[e.g.,][]{kp06}.
Isolated stars should have smaller (similar) masses compared with the case of NS-WD (NS-NS) binaries, respectively. The long-standing accretion from companions make the primaries heavier in the case of NS-WD binary systems~\citep{bog05}.
Although a single EoS must be applied to all the compact stars, the existing phase of matter depends on the density.
Therefore, the location of the Cas A observation on the $(T_{\rm eff}-t)$ plane
 becomes very difficult to interpret if we
believe the models with strong cooling mechanisms explain all other the observations of $T_{\rm eff}$.

In this paper,
we present models that satisfy both cases of Cas A and other cooled stars such as 3C58 and/or Vela, by considering hybrid stars composed of quark matter, hadron matter, and their mixed phase (MP), where a characteristic property of
color superconducting (CSC) phase is utilized. In addition, we also show cooling curves of Cas A for observations over the past 10 years and indicate that the phase transition to the superfluidity is consistent with the observations.

\section{Cooling curve models}

We construct a model that includes both quark--hadron MP and its CSC phase. Considering the 
first-order phase transition between hadron and quark phases, it would be plausible that both phases coexist and form some kind of MP. Similar to the ``nuclear pasta'' phase in the crust of a neutron star~\citep{rev83,hashi84}, it has been shown that MP could form geometrical structures~\citep{maru08}; \citet{yasu09} have made EoS of an MP under a Wigner-Seitz (hereafter ``WS'') approximation using an MIT Bag model for a quark phase in finite temperature. In the present study,
we employ an EoS with the same framework using the bag constant
$B=100~ \mathrm{MeV~fm^{-3}}$, the coupling constant $\alpha_\mathrm{S}=0.2$, and the surface tension parameter $\sigma = 40~\mathrm{MeV~fm^{-2}}$.
For hadron phase, we adopt the results of the Bruekner--Hartree--Fock (BHF) theory including hyperons, $\Lambda$, and $\Sigma^-$~\citep{shlz95, bald98, bald99}.
However, the hyperons do not appear for the EoS calculation including geometrically structured MP~\citep{yasu09}; therefore, we do not include the effects of hyperons. Although this does not occur in our model, if hyperons appear in other models, the hyperon mixed matter has large neutrino emissivity called hyperon direct URCA process (e.g., $\Lambda \to p + e^- + \bar{\nu_e}$), and causes the rapid cooling of compact stars~\citep{tak97}.
Since the BHF results are inappropriate for low-density matter in
the crust, 
 we apply  EoS of BPS~\citep{bps71} for the crust.
The EoS gives a maximum mass of $1.53M_\odot$ with a radius of $8.6~ \mathrm{km}$, and the mass lies within the limits of the observation of Cas A.
Although our EoS is inconsistent with the recent observation of the mass $M \sim 2 M_\odot$ of pulsar J1614--2230~\citep{dem10}, we could overcome this issue by adopting other EoS models~\citep[e.g.,][]{alf05}. 

\begin{figure}[t]
	\begin{center}
		\includegraphics[width=1.0\linewidth,keepaspectratio,clip]{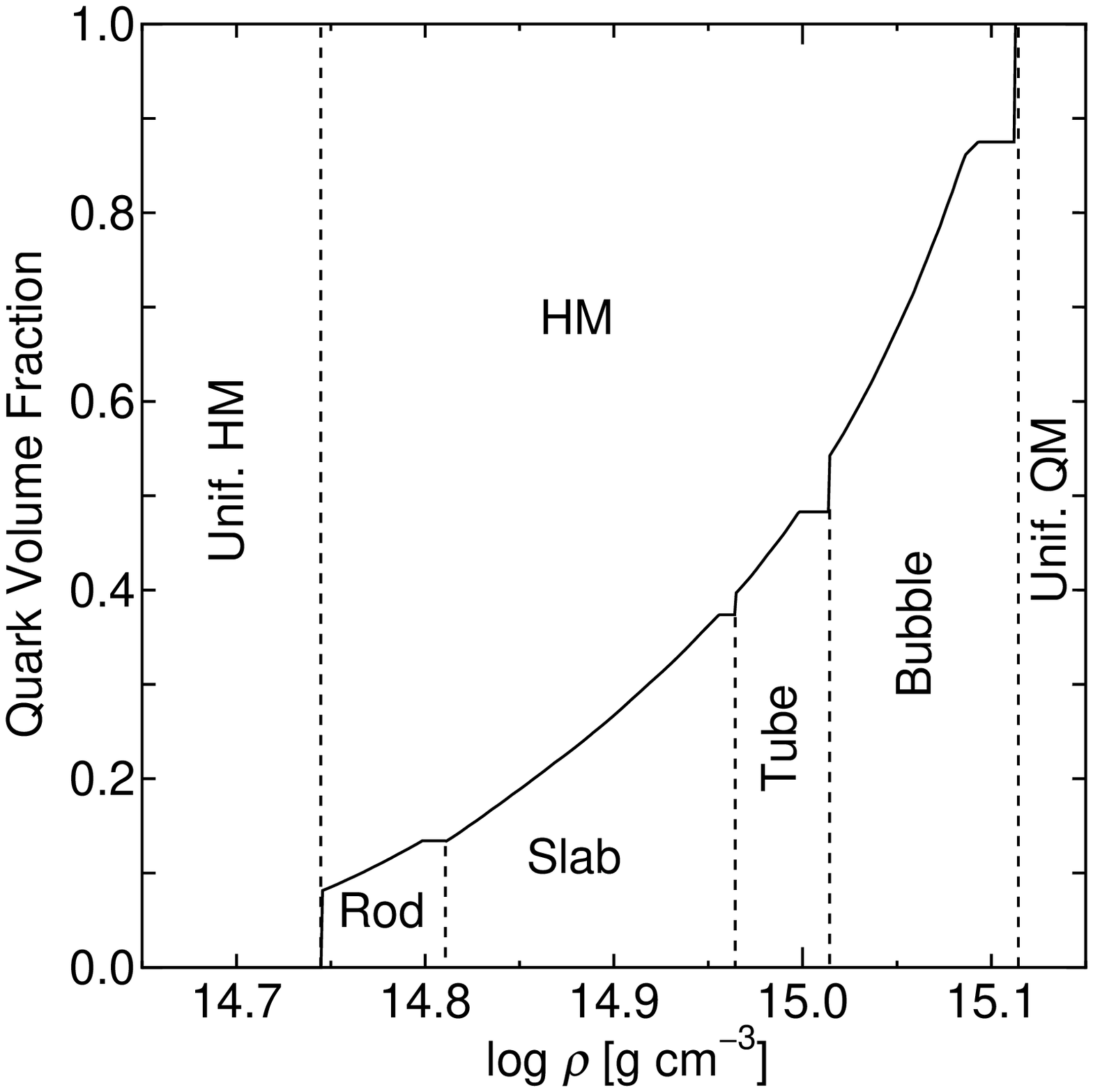}
	\end{center}
	\caption{Volume fractions of quark matter phase having particular geometrical structures with a bag constant $B=100~\mathrm{MeV}~\mathrm{fm}^{-3}$ and a coupling constant
 $\alpha_\mathrm{s} = 0.2$~\citep{maru07,yasu09}.}
	\label{fig:MP}
\end{figure}

Using the WS approximation, we obtain a cell radius of each phase and calculate the volume fraction of quark matter in MP as seen in Figure \ref{fig:MP}.
It is difficult to calculate the neutrino emissivity in MP. Therefore,
the volume fraction $F$ is multiplied by
the original quark neutrino emissivity $\varepsilon_{\nu,0}$~\citep{iwa80}; the total emissivity by quarks is set to be $\varepsilon_\nu = F \varepsilon_{\nu,0}$.
We adopt well-known neutrino emission processes without nucleon superfluidity for hadronic matter~\citep{fri79}: modified URCA process for the higher density region and Bremsstrahlung process for the crust. We note that the special case of Cas A during the past 10 years is discussed in Section 3.

The color superconductivity is the key of the present study. There are different kinds of quark pairings such as CFL (Color Flavor Locking) or 2SC (Two-Flavor Color Superconducting) according to the degrees of freedom of quark flavor and color. It is considered that the 
energy gap of $\Delta \gtrsim 10~\mathrm{MeV}$ is very large compared with the temperature of the center of compact stars , $T_\mathrm{C} \sim \mathrm{keV}$~\citep[e.g.,][]{shm10}.
Once matter becomes superconducting, neutrino emissivity must be suppressed due to the large energy
gap
and it could be proportional to $\exp\left(-\Delta/k_\mathrm{B}T\right)$, where $T$ is the temperature at the relevant layer and $k_{\rm B}$ is the Boltzmann constant~\citep{neg12}. Therefore, in the CSC phase with a large energy gap ($\Delta \gg T$), neutrino emissivity by quarks is almost negligible~\citep{alf08}.
We note that if we adopt a large energy gap we do not need to consider which kinds of CSC pairing appear.
In particular, the difference between CFL and 2SC is unimportant for the cooling. In fact, the pseudo NG bosons may exist and decay in the CFL phase and make changes to neutrino emissivity~\citep{jkm02}. However, it is known that it is not quantitatively affected in low temperature region~\citep{alf08}.

We assume that the energy gap in the CSC phase is very large and the phase appears in the
density region above the threshold density $\rho_\mathrm{T}$. For simplicity, we use the critical volume fraction $F_\mathrm{C}$ instead of $\rho_\mathrm{T}$. If the matter has a volume fraction $F > F_\mathrm{C}$ at the density in the layer of MP, quarks change into a CSC state. 
As a consequence, the neutrino emission process due to the quark $\beta$ decay works only in the lower density region of the quark--hadron MP.

\begin{figure}[!h]
	\begin{center}
		\includegraphics[width=1.0\linewidth,keepaspectratio,clip]{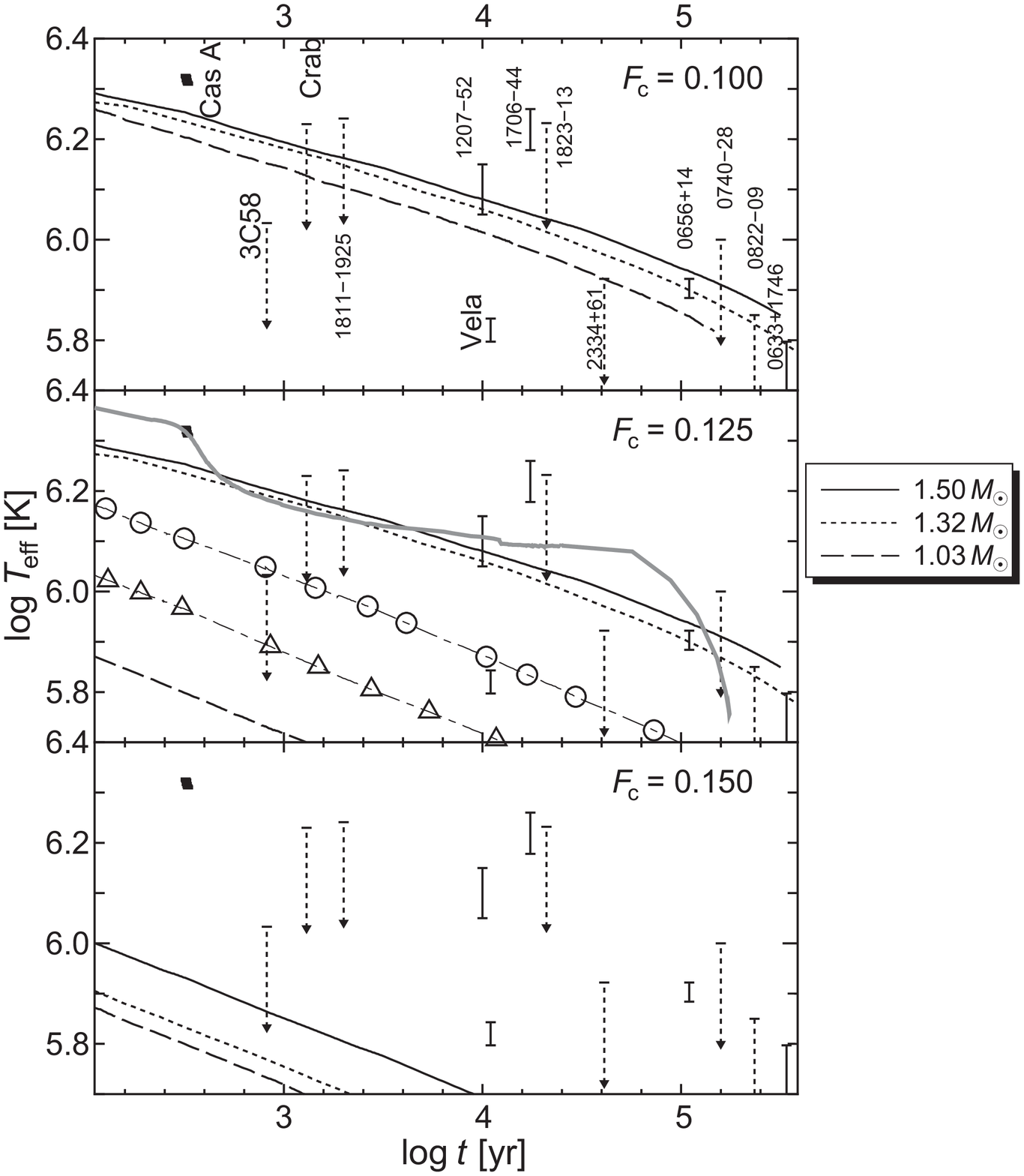}
	\end{center}
	\caption{Cooling curves with color superconducting quark phases. Solid, dotted, and dashed lines denote the models with the masses of $1.50$, $1.32$, and $1.03 ~M_\odot$, respectively. Thick gray line on the middle panel denotes a $1.50M_\odot$ model with nucleon superfluidity and carbon envelope. Dot-dashed lines with marks on the middle panel indicate the model of the mass $1.03 M_\odot$ except for the neutrino emissivity in normal quark phase multiplied by one-tenth and one-one-hundredth for the lines with
marks of triangle and circle, respectively.}
	\label{fig:CC}
\end{figure}

We select stellar masses of compact stars to be
$1.0$, $1.3$, and $1.5 M_\odot$ whose central densities are 1.48, 1.82, and 
2.67~$\times10^{15}~\rm g~cm^{-3}$, respectively.
We take the critical fraction $F_\mathrm{C}$ to be 0.1, 0.125,
and 0.2
which are appropriate to explain the observations with our scenario.
The results are shown in Figure \ref{fig:CC}, 
with available observational values.
Most of the observational data are taken from the Table 7.2 in~\citet{kp06}:~\citet{sl02} for 3C58;~\citet{wk04} for Crab;~\citet{pl01} for Vela;~\citet{bk03} for 0656+14;~\citet{gh02} for 1706--44;~\citet{kp06} 
for 1811--1925;~\citet{gs03} for 1823--13;~\citet{bk96} for 2334+61. Other data are taken from~\citet{pg02}. The data of Cas A are attached~\citep{ho10} as the youngest compact star.
We find that the cooling curves transit from hotter regions to cooler regions 
for the parameter between $0.1 < F_\mathrm{C} < 0.2$. As indicated in the middle panel of Figure \ref{fig:CC}, the cooling curves split into two regions for $F_\mathrm{C}=0.125$ and stars with larger masses cool more slowly than those with lighter masses. Since lighter mass stars cool faster, they are suitable for the 3C58 case which does not have lower limit of the effective temperature observation. However, the calculated cooling curves are inconsistent with the Vela case which has lower limit. Also, the quark cooling is still too strong to explain this case.

Since the neutrino emissivity of a quark phase involves large uncertainty, we have calculated the additional cooling curves for the mass $1.03 M_\odot$ in the case of $F_C=0.125$ with the neutrino emissivity reduced by a factor of $0.1$ and $0.01$.
There are some possible factors of this reduction for neutrino emissivity accompanying quark $\beta$-decay, such as an increase of the abundance of strange quarks; a decrease in electron numbers inside MP leads to a reduction of neutrino emissivity \citep{iwa80}. The presence of 2SC at low density also suppresses the emissivity; \citet{maru08} discussed that the abundance of quarks in MP changes and may cause CSC phase. We suppose that the reduction of emissivity originates from the above physical processes.
If the emissivity of quarks is reduced by these factors, the observation of Vela can be explained as shown
in the middle panel of  Figure \ref{fig:CC}.

\section{Discussions}

We demonstrate the effect of color superconductivity in quark--hadron MP on the cooling
curve: the larger the masses of the compact stars, the slower the speed of the cooling.
This situation is caused by the layer that emits a large number of neutrinos through quark $\beta$ decay processes, which encircles the center of the star. The thickness of this layer decreases as stellar mass increases.
Although the maximum mass of our model is at most $1.53~M_{\odot}$ that would be near the lower limit
for the recent observation of Cas A, our cooling scenario can be applied 
for the cases of $M > 1.5 M_{\odot}$ if more concrete EoS are devised.
Some problems also remain concerning the fundamental physics: uncertain physical properties of quark--hadron MP, indefinite tuning of the threshold density of CSC, unknown values of the energy gap $\Delta$. Nonetheless, to explain the observed effective temperature of Cas A, our model incorporating the color superconductivity with a large energy gap is compatible with available observational data. The cooling mechanism associated with CSC quarks could be plausible because, as shown by the existence of CSC, it may be quite natural from the recent study of phase diagram between quarks and hadrons~\citep{rus05}.

The fundamental physics associated with compact stars is still largely uncertain.
In the quark phase, the abundance of each quark is still unknown, and therefore the neutrino emissivity is not determined from the fractions of $u$, $d$, and $s$ quarks and/or the chemical potential of electrons~\citep{maru08}.
Although the existence of the CSC phase has been well studied, it is still open to debate which type of CSC appears, and the quantitative values of the critical density and the energy gap should be clarified.
Since the physical properties of quarks in the MP are different from those in the uniform phase, there would be many factors to change the emissivity.
Further theoretical study in collaboration with the observations is required to constrain the physics of compact stars.

\begin{figure}[!h]
	\begin{center}
		\includegraphics[width=0.48\linewidth,keepaspectratio,clip]{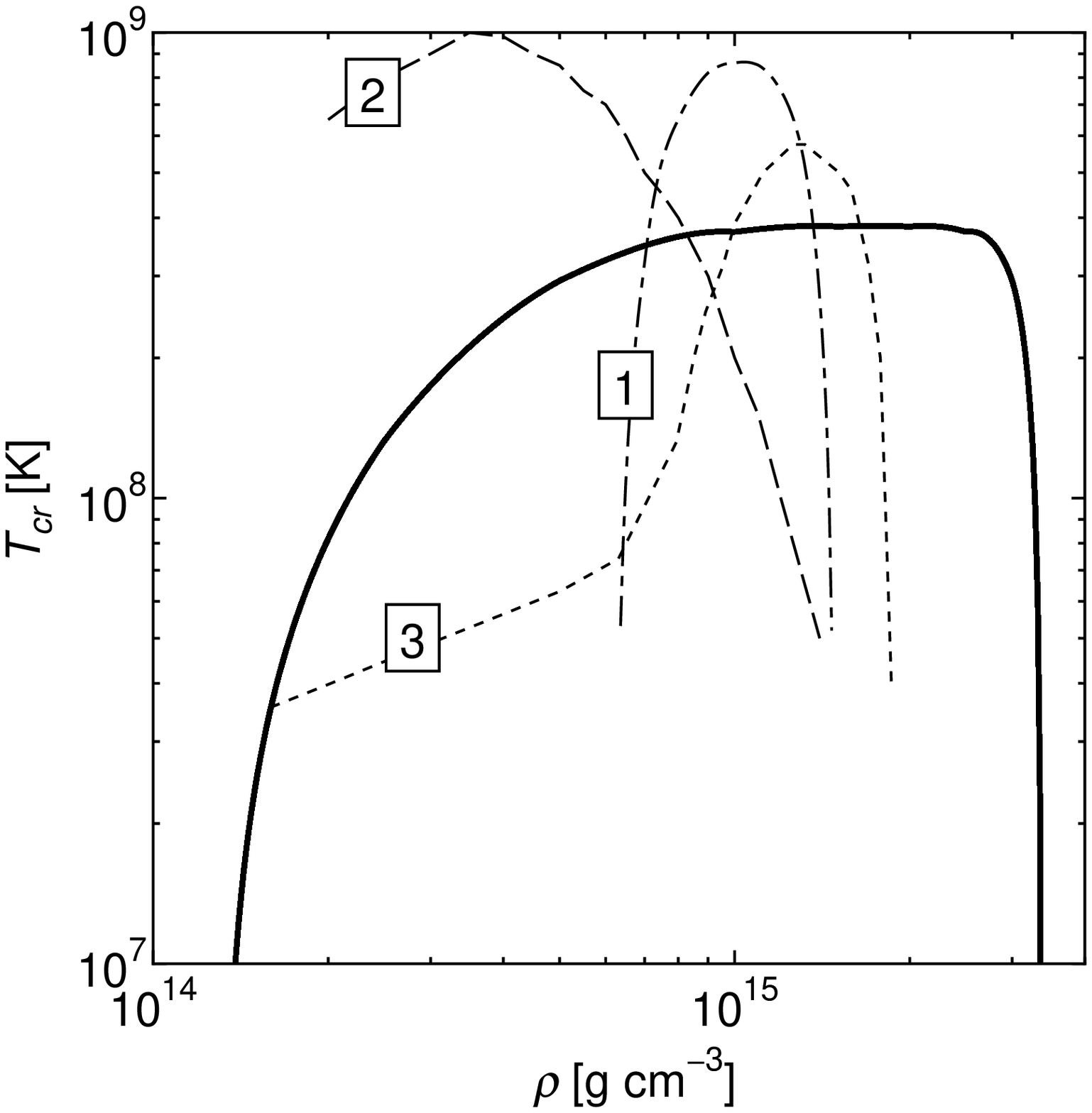}
		\includegraphics[width=0.48\linewidth,keepaspectratio,clip]{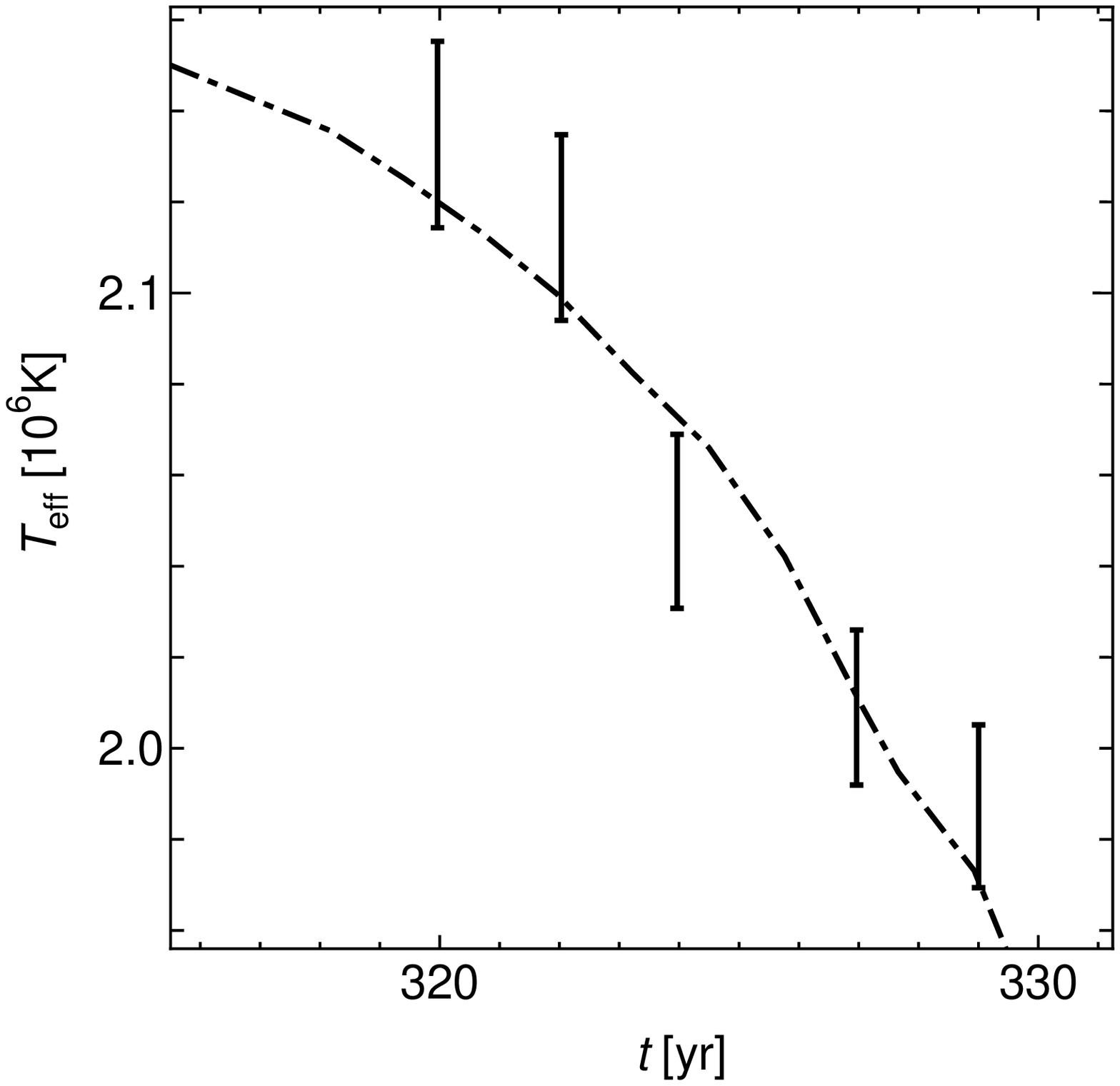}
	\end{center}
	\caption{Left panel: density dependence of the critical temperature for neutron ${}^3P_2$ superfluidity used by this study (solid and thick line) and previous works. Dot-dashed (labeled ``1''), long dashed (labeled ``2''), and short dashed (labeled ``3'') denote the critical temperature adopted by \citet{sht11}, \citet{pg09}, and \citet{kam06}, respectively.
	Right panel: cooling curve with the nucleon superfluidity which crosses the observational data of $T_\mathrm{eff}$ in Cas A~\citep{ho10}.}

	\label{fig:NSCCC}
\end{figure}

%%%%
Considering the above uncertainties, the observations of the rapid cooling for Cas A ~\citep{ho10}
would give insight for constraining some properties of high density matter.
We have tried to fit the observational data using a model with nucleon superfluidity.
We adopt the neutrino emissivity accompanied by the phase transition from the normal state to that of
the nucleon superfluidity (\citet{yak04}), where we tune the critical temperature of the neutron ${}^3P_2$ superfluidity and the associated neutrino emissivity.
%%%
For simplicity, we adopted only the superfluidity effect of the neutron ${}^3P_2$, not of the neutron/proton ${}^1S_0$. The singlets affect the cooling of a compact star, but the neutron singlet works in a lower density than the triplet, and the proton singlet is still speculative. The most effective nucleon superfluidity is caused by the neutron triplet state~\citep{pg11b}.
%%%
Since ~\citet{ho09} concluded that in order to reproduce the observations in the X-ray spectrum the surface composition of Cas A must be carbon and/or helium, we set the surface composition to be of carbon and a small amount of helium.
This is because the
existence of carbon results in a rather high effective temperature at the beginning of cooling phase.
To calculate the cooling curves,
we assume the functional form of the critical temperature which is
a phenomenological extension of  that for the ${}^3P_2$ neutron superfluidity as seen in the left panel of Figure \ref{fig:NSCCC}.
This approach is similar to the method used by \citet{sht11} except for the profile of the critical temperature.
Considering that the cooling curve sensitively depends on the critical temperature (left panel in Figure \ref{fig:NSCCC}) and the neutrino emissivity, we have fine tuned the two quantities to fit the observational data($T_\mathrm{eff}$) of Cas A over the past 10 years (right panel in Figure \ref{fig:NSCCC}).

%%%%

There are some experimental projects of hadron colliders  with intermediate energy, such as J-PARC or GSI, that may help us to understand the state of ultrahigh density. They are very useful for examining high density character composed of hyperons, mesons, transition to  normal quark matter and CSC phase \citep{andr10}. However, it is still difficult to reproduce the same phase as in the  core of compact stars by the colliders, where they produce high temperature ($T\sim10$~MeV) in the high density region of $\rho\sim10^{15} \mathrm{gcm^{-3}}$ compared
 with the core of compact stars \citep{andr09}. Therefore, it is worthwhile to check the theories by both observing as many compact stars as possible and comparing theoretical predictions with observations such as the effective temperature.
From this viewpoint, further observations of Cas A and central sources of other supernova remnants such as SN1987A are necessary to understand the fundamental physics in these extreme conditions.

%%%%
There are some recent studies of hybrid star cooling, such as \citet{neg12}, \citet{yin11} or \citet{schr12}. \citet{neg12} employed a similar model to our model, but the assumed energy gap $\Delta$ is in the range of $0.1$---$1.0~\mathrm{MeV}$, and the pair breaking and formation (PBF) process is not included; thus resulting rapid decrease in the effective temperature of Cas A is not compatible with their model.  \citet{yin11} considered the quark--hadron mixture and direct URCA process in hadronic phase, but did not included quark CSC, and resulting the cooling with the direct URCA is too strong to explain the observational data of older compact stars. 
\citet{schr12} adopted the rotation of compact stars which delays the isothermal relaxation, and this effect would help us to understand the temperature drop of Cas A. These effects should be included in our further study.
%%%%

Even if the quark--hadron MP does not exist, our scenario of color superconducting core surrounded by exotic phase could be applied to a cooled object such as Vela.
Considering a meson condensed phase, a nucleon superfluidity, which reduces the strong neutrino emission by mesons, is expected to explain the Vela data consistent with cooling curves with some mass ranges.
The cooling scenario of compact stars not only affects the cooling of isolated stars, but also the stars in binary systems. There are some observational data of X-ray transients~\citep[e.g.,][]{rut02} that have gravitational energy supply on the surface due to accretion from companions, and X-ray bursts could result from thermonuclear reactions on the surface. These systems are worthwhile to re-examine from the point of view of exotic cooling as was partly done in \citet{yak04} for a quiescence period.
Since binary systems have larger luminosity and/or orbital information such as inclination angle, rotational
period, and a signal of the gravitational wave, more accurate mass detection than isolated compact stars would be possible~\citep[e.g.,][]{kra06, wei10, lat10}.

This work has been supported in part by the Grant-in-Aid for Scientific Research (19104006, 21105512, 21540272, and 24540278) and the Grant-in-Aid for the Global COE Program ``The Next Generation of Physics, Spun from Universality and Emergence'' from the Ministry of Education, Culture, Sports, Science and Technology (MEXT) of Japan.

\end{document}